\documentclass[11pt,a4paper]{article}

\usepackage{latexsym}
\usepackage{amsmath}
\usepackage{epsfig}
\usepackage{times}
\usepackage{amssymb}

\newtheorem{thr}{\quad Theorem}
\newtheorem{lem}{\quad Lemma}
\newtheorem{cor}{\quad Corollary}
\newtheorem{defin}{\quad Definition}
\newtheorem{que}{\quad Question}

\newcommand{\F}{\mathbb{F}}
\newcommand{\Q}{\mathbb{Q}}

\date{}

\title{Upper Bounds on the Number of Codewords of Some Separating Codes}

\author{*Ryul Kim, Myong-Son Sin, Ok-Hyon Song \\ \\
\small{Faculty of Mathematics, \textbf{Kim Il Sung} University,  D.P.R.Korea } \\
\small{*e-mail: ryul$\_$kim@yahoo.com}}

\begin{document}
\maketitle

\begin{abstract}
Separating codes have their applications in collusion-secure fingerprinting for generic digital data,
while they are also related to the other structures including hash family, intersection code and
group testing. In this paper we study upper bounds for separating codes. First, some new upper bound
for restricted  separating codes is proposed. Then we illustrate that the Upper Bound Conjecture for
separating Reed-Solomon codes inherited from Silverberg's question holds true for almost all Reed-Solomon
codes.

\bigskip

\end{abstract}

%
%
\section{Introduction}

Let $\Q$ be an arbitrary set of $q$ elements, $n$ be a positive integer, and $C$ be a code of length
$n$ with the alphabet set $\Q$. For a nonempty subset $U$ of $C$ we define \textit{descendant set} and
\textit{feasible set} by $\textnormal{desc} U:=\{x \in \Q^n |$ $\textnormal{ for every } i
\textnormal{ there exists } a \in U \textnormal{ such that } a_i=x_i\}$ and $F(U):=\{x \in \Q^n |$
if all words in $U$ coincide on $i$th coordinate for some $i$, then $x_i$ also takes the value.$\}$,
respectively, where $x_i$ denotes the $i$th coordinate of vector $x$.

%
%
\begin{defin}
Let $w_1, w_2$ be positive integers and let's assume that at least one of them is larger than one.
The code $C$ is said to be $(w_1, w_2)-$separating code, if the descendant sets of any two disjoint
subsets of $C$ with not more than $w_1$ and $w_2$ codewords, respectively, are also disjoint.
By replacing descendant sets by feasible sets, we get the definition of restricted $(w_1, w_2)-$
separating codes.
\end{defin}

We call $(w, 1)-$separating code by $w-$FP code, and $(w, w)-$separating code by $w-$SFP code for $w>1$.
Since separating codes are powerful weapon of anti-collusion fingerprinting, many recent works were
done in the literatures, e.g., \cite{boneh}.  Particularly, the upper bound on the number of codewords in separating
codes for given alphabet size $q$ and code length $n$ has been considered.
The strongest upper bound ever found for $w-$SFP codes is 
$M \leq (2w^2-3w+2)q^{\lceil\frac{n}{2w-1} \rceil}-2w^2+3w-1$ of \cite{stins}, where the result for
$(w_1, w_2)-$separating codes were also suggested.
Restricted separating codes were introduced in \cite{pin}, and their behaviors such as
the bound of code rate were investigated in \cite{barg, sag} and so on. They have still
wider application than separating codes, although their upper bound has not been
studied in earlier works.
To understand Silverberg's conjecture and related upper bound question, we need to refer to
the concept of IPP code.

%
%
\begin{defin}
Let $C$ be a code of length $n$ and $w \geq 2$ be a positive integer. The code $C$ is said to be
$w-$IPP, if for any $x \in \Q^n$, the intersection of all subsets of $C$ that contain
not more than $w$ codewords and involve $x$ in the corresponding descendant set, is not
empty.
\end{defin}

IPP(\textit{Identifiable Parent Property}) code is another important class of fingerprinting
codes. It is easy to prove that $w-$IPP implies $w-$SFP. The following results are well
known in fingerprinting code theory.

%
%
\begin{thr}
(Theorem 4.4 in \cite{stad}) Let $C$ be a code of length $n$. If the minimum distance of $C$
satisfies $d>n(1-\frac{1}{w^2})$, then $C$ is a $w-$IPP code.
\end{thr}

%
%
\begin{thr}
(Proposition 7 in \cite{coh}) Let $C$ be a code of length $n$. If the minimum distance of $C$
satisfies $d>n(1-\frac{1}{w_1 w_2})$, then $C$ is a $(w_1, w_2)-$separating code.
\end{thr}

In \cite{sil}, Silverberg considered applications of Reed-Solomon codes as well as other algebraic
geometry codes to collusion-secure fingerprinting techniques, where he proposed the following
open problem.

%
%
\begin{que}
Is it the case that all $w-$IPP Reed-Solomon codes satisfy the condition $d>n(1-\frac{1}{w^2})$?
\end{que}

For Reed-Solomon codes, $d=n-k+1=q-k$ so we can replace the statement $d>n(1-\frac{1}{w^2})$ with
$k< \frac{q-1}{w^2}+1$. Since the number of codewords in Reed-Solomon code of dimension $k$ is
$M=q^k$, it now equals with $M \leq q^ {\lceil \frac{n}{w^2} \rceil}$. Thus, Silverberg's problem conjectures
the upper bound of IPP Reed-Solomon codes, which is exactly optimal if true from Theorem 1.
Silverberg's problem was studied in \cite{fer}. They showed that a large family of Reed-Solomon codes
holds Question 1 positive. What is interesting for their work is that the family satisfies more general
fact. The main result of \cite{fer} is as follows. From now we denote Reed-Solomon code of dimension $k$
over $\F_q$ by $RS_k(q)$.

%
%
\begin{thr}
(Theorem 7 in \cite{fer}) Suppose that ${k-1} \mid {q-1}$. If the code $RS_k(q)$ is $(w_1, w_2)-$
separating, then $k<\frac{q-1}{w_1 w_2}+1$.
\end{thr}

We can easily check that Theorem 3 suggests the conjecture of the upper bound
$M \leq q^ {\lceil \frac{n}{w_1 w_2} \rceil}$
for separating Reed-Solomon codes.

%
%
\begin{que}
(Upper Bound Conjecture for Separating Reed-Solomon Codes) Is it the case that all $(w_1, w_2)-$separating
Reed-Solomon codes satisfy the condition $d>n(1-\frac{1}{w_1 w_2})$?
\end{que}

If Question 2 holds positive for all cases, then it would turn out we obtain the optimal upper bound of separating
Reed-Solomon codes by Theorem 2. The proof of that, however, is not easy.
The goal of this paper is firstly, to get a new upper bound for restricted  separating codes, and secondly to
illustrate that almost all separating Reed-Solomon codes involving those of \cite{fer} allow the positive
answer for Question 2.

%
%
\section{Main Results}

%
%

\subsection{Upper Bound for Restricted Separating Codes}

Our new bound for restricted $(w,w)-$separating code is stated in Theorem 4. Note that the bound
is independent on alphabet size $q$.

%
%
\begin{thr}
Let $w \geq 3$ be a positive integer. If $C$ is a code of length $n$ with $M$ codewords and satisfies restricted
$(w,w)-$separation property, then
\begin{equation*}
M \leq 2^{\lfloor \frac{n-w+2}{2}  \rfloor}+w-2
\end{equation*}
\end{thr}

\textit{Proof.} Pick an arbitrary subset $U$ of $C$ with $w-2$ codewords. We can assume that all the
elements of $U=\{x^{(1)}, \cdots, x^{(w-2)}\}$ coincide on and only on the first $d$ coordinates.
Set $S=\{1,2, \cdots, d\}$ and define $\Gamma (y):=\{i \in S \mid y_i = x^{(1)}_i\}$ for all
$y \in C \backslash U$. If $y, z, t \in C \backslash U$ are distinct elements, then the followings hold true.

\begin{eqnarray*}
&& (1) \; \Gamma (y) \cap \Gamma (z) \neq \varnothing \\
&& (2) \; \Gamma (y) \not{\subset} \Gamma (z) \\
&& (3) \; \Gamma (y) \cap \Gamma (z) \neq S \\
&& (4) \; \Gamma (y) \cap \Gamma (z) \not{\subset} \Gamma (t) \\
&& (5) \; \Gamma (t) \not{\subset} \Gamma (y) \cup \Gamma (z),
\end{eqnarray*}

\noindent
since the negations imply $F(U \cup \{y,z\})=\Q^n, F(U \cup \{y\}) \cap F(\{z\})=\{z\}$,
$F(U) \cap F(\{y,z\}) \neq \varnothing, F(U \cup \{y,z\}) \cap F(\{t\})=\{t\}$ and
$F(U \cup \{t\}) \cap F(\{y,z\}) \neq \varnothing$, respectively, that all contradict the restricted 
$(w,w)-$ separation property of $C$.
\\ \indent
\textit{Case 1:} Assume that there exists $y^{(0)} \in C \backslash U$ such that
$|\Gamma (y^{(0)})| \leq \lfloor \frac{d}{2} \rfloor$. For all $y \in C \backslash U$, define
the correspondence $\Gamma' (y) := \Gamma (y) \cap \Gamma (y^{(0)})$. Then
$\Gamma'$ is an injection from (4). For $\Gamma'$ maps $C \backslash U$ to
$\Gamma (y^{(0)})$ of at most $ \lfloor \frac{d}{2} \rfloor$ elements, we get
$|C \backslash U| \leq 2^{ \lfloor \frac{d}{2} \rfloor}$.
\\ \indent
\textit{Case 2:} Assume that for all $y \in C \backslash U$, $|\Gamma (y)| >  \lfloor \frac{d}{2} \rfloor$.
Set $\Gamma _1 (y) := S \backslash \Gamma (y)$, then $\Gamma _1$ also satisfies (1)-(5).
Similarly as above, we get $|C \backslash U| \leq 2^{ \lfloor \frac{d}{2} \rfloor}$.
\\ \indent
From the definition of restricted separating code, we directly get $d \leq n-w+2$.
Combining two results above, $|C|=|U|+|C \backslash U| \leq 2^{\lfloor \frac{n-w+2}{2} \rfloor}+w-2$. $\Box$

%
%

\subsection{Optimal Upper Bound for Separating Reed-Solomon Codes}

In the previous section we obtained new upper bounds for some separating codes. This section, however,
is a little different. We are dealing with separating codes included in Reed-Solomon codes family
and are proving the Upper Bound Conjecture derived from Silverberg's problem, which is to be optimal.
Let $\F_q$ be a finite field of characteristic $p$ with a primitive element $\alpha$. Denote the set
of all non-zero polynomials over $\F_q$ of degree less than $k$ by $P_k$. The following lemma is trivial
from definition so that we are going to state without proof.

%
%
\begin{lem}
Assume that $RS_k(q)$ is not $(w_1,w_2)-$separating, then
\begin{eqnarray*}
&& (1) \; q-1 \geq l \geq k \textnormal{ implies that $RS_l(q)$ is not $(w_1,w_2)-$separating}. \\
&& (2) \; w'_1 \geq w_1, w'_2 \geq w_2 \textnormal{ implies that $RS_k(q)$ is not $(w'_1,w'_2)-$separating}.
\end{eqnarray*}
\end{lem}

In \cite{fer}, they gave the equivalent condition with separation property of Reed-Solomon codes before
they evolved the relation between $k$ and $q$, namely, ${k-1} \mid {q-1}$. Similarly, we state the
following sufficient condition for non-separation of Reed-Solomon codes at first.

%
%
\begin{lem}
Let $f$ be a non-constant polynomial belonging to $P_k$. Suppose there exist two subsets $E, F$ of
Im$f$ such that $1 \leq |E| \leq w_1, \: 1 \leq |F| \leq w_2$ and either of the two facts
Im$f=EF$ or Im$f=E+F$ holds true. Then, the code $C=RS_k(q)$ is not $(w_1,w_2)-$separating.
\end{lem}

\textit{Proof.} We will show only in the case Im$f=E+F$, since the other case can be proven similarly.
Define $U:=\{ev(\beta) \mid \beta \in E \}$ and $V:=\{ev(f-\gamma) \mid \gamma \in F \}$. $U, V$ are
nonempty sets of at most $w_1, w_2$ elements, respectively. Further, they are disjoint since $f$ is
non-constant.
For all $i(1 \leq i \leq q-1)$, there exist $\beta_i \in E, \gamma_i \in F$ such that
$f(\alpha^i)=\beta_i+\gamma_i \in \textnormal{Im} f$ since $\alpha^i \in \F_q$. Set $x:=(\beta_1, \cdots, \beta_{q-1})$,
then we can easily check that $x$ belongs to desc$U \cap$desc$V$. Therefore,
$C=RS_k(q)$ is not $(w_1,w_2)-$separating. $\Box$

\smallskip
Lemma 2 allows us to discuss the relation between $k, q, w_1, w_2$ that are parameters specifying
separation property and Reed-Solomon codes to meet the positive answer for Question 2. First, we give
a different proof of Theorem 3 using Lemma 2 to show generality of our results.

\smallskip
\textit{Proof of Theorem 3.} Assume $k \geq \frac{q-1}{w_1 w_2}+1$ and define $f(x):=x^{k-1}$.
Then $f$ is a polynomial of $P_k$ and it is a multiplicative homomorphism over $\F^*_q$. Therefore
Im$f$ is a subgroup of $\F^*_q$, and thus, is cyclic. Let $\gamma$ be a generator of Im$f$, and
set $E:=\{\gamma ^ {iw_2} \mid 0 \leq i \leq w_1-1 \}, \: F:=\{\gamma ^ j \mid 0 \leq j \leq w_2-1\}$.
Applying group theory, we get $| \textnormal{Im} f|=\frac{q-1}{k-1} \leq w_1 w_2$ and Im$f=EF$
since $|$Ker$f|=k-1$. Thus, the conditions of Lemma 2 satisfy and $RS_k(q)$ is not $(w_1,w_2)-$separating. $\Box$

\smallskip
Here we are to find new relation of parameters for satisfying Upper Bound Conjecture in terms of Lemma 2.
Let $r_1:=[\log {_p}{w_1}], \: r_2:=[\log {_p}{w_2}]$.

%
%
\begin{thr}
Suppose ${k-1} \mid q$ and at least one of the following conditions is true.
\begin{eqnarray*}
&& (1) \; k-1 \geq \frac{p q}{w_1 w_2} \\
&& (2) \; \frac{w_1}{p^{r_1}} \cdot \frac{w_2}{p^{r_2}}<p \\
&& (3) \; [\frac{w_1}{p^{r_1}}] \cdot [\frac{w_2}{p^{r_2}}] \geq p
\end{eqnarray*}
If $RS_k(q)$ is $(w_1,w_2)-$separating, then $k < \frac{q-1}{w_1 w_2} + 1$.
\end{thr}

\textit{Proof.} Set $s:=k-1$ for convenience and assume $s \geq \frac{q-1}{w_1 w_2}$ in spite that
$RS_k(q)$ is  $(w_1,w_2)-$separating. Define $f(x):=x^s-x$. Since the characteristic of the field is $p$ and $s$ is
a power of $p$, $f$ is an additive homomorphism from $\F_q$ to $\F_q$ and its kernel is Ker$f=\F_s$,
therefore $|$Im$f|=q/s$. \par
Assume (1) is true. Then $|$Im$f|=q/s \leq \frac{w_1 w_2}{p} \leq p^{r_1+r_2}$. For $|$Im$f|$ is a power of
$p$, there exist $t_1, t_2(t_1 \leq r_1, t_2 \leq r_2)$ such that $|$Im$f|=p^{t_1+t_2}$ . According to
group theory, there exist subgroups $E$ and $F$ of Im$f$ such that $|E|=p^{t_1} \leq w_1, |F|=p^{t_2} \leq w_2$,
and Im$f=E+F$. Applying Lemma 2 leads to the contradiction to $(w_1,w_2)-$separation property. \par
Assume that (2) is true. Then we get $|$Im$f|=q/s \leq w_1 w_2 < p^{r_1+r_2+1}$ and since $|$Im$f|$ is
a power of $p$, it equals with $|$Im$f| \leq p^{r_1+r_2}$. So the exactly same discussion as above holds
in this case. \par
Finally, assume that (1), (2) is false but (3) is true. Failure of (1) implies the fact
$\frac{q}{w_1 w_2} \leq s \leq \frac{p q}{w_1 w_2}$, and the equality can not be held in (3) for $p$ is
a prime number. Thus, $w_1 w_2>p^{r_1+r_2}$ . If we consider $p^{r_1+r_2+2}>w_1 w_2$, we get the series
of inequalities such as $p^{r_1+r_2} < \frac{w_1 w_2}{p} <|$Im$f|=q/s \leq w_1 w_2 <p^{r_1+r_2+2}$.
So $|$Im$f|=p^{r_1+r_2+1}$ since $|$Im$f|$ is a power of $p$. Then there exist subgroups $E', F', P$
of Im$f$ such that Im$f=E'+F'+P$ and their orders are $p^r_1, p^r_2$, and $p$, respectively. Moreover,
$P$ is cyclic as its order is a prime number. Denote one of the generators of $P$ by $\gamma$ and set
$P_1:=\{i[\frac{c_2}{p^{r_2}}] \gamma \mid 0 \leq i \leq [\frac{c_1}{p^{r_1}}]-1\}$,
$P_2:=\{j \gamma \mid 0 \leq j \leq [\frac{c_2}{p^{r_2}}]-1\}$. Then $P=P_1+P_2$ since
$[\frac{c_1}{p^{r_1}}] \cdot [\frac{c_2}{p^{r_2}}] \geq p$.
Now let $E:=E'+P_1, \: F:=F'+P_2$. The sizes of $E, F$ are $p^{r_1} \cdot [\frac{c_1}{p^{r_1}}]$ and
$p^{r_2} \cdot [\frac{c_2}{p^{r_2}}]$, respectively, so $1 \leq |E| \leq c_1, 1 \leq |F| \leq c_2$ and
Im$f=E+F$. Therefore, we get contradiction to the separation property of $RS_k(q)$ applying Lemma 2. \par
Thus, the statement of the theorem holds true in all cases. $\Box$

\smallskip
If for some $k$ we know that $(w_1, w_2)-$ separation property of $RS_k(q)$ implies
$k < \frac{q-1}{w_1 w_2} + 1$, then for all integers larger than $k$ the same holds true by Lemma 1.
It inspired us to believe that all Reed-Solomon codes employ the conjecture. \par
The following corollaries are simple to prove.

%
%
\begin{cor}
Suppose that $w_1 w_2 \geq q-1$ or $w_1 w_2 \mid q-1$. If the code $RS_k(q)$ is $(w_1, w_2)-$separating, then
$k < \frac{q-1}{w_1 w_2} + 1$.
\end{cor}

%
%
\begin{cor}
Suppose $w_1 w_2 \mid q$. If the code $RS_k(q)$ is $(w_1, w_2)-$separating, then $k < \frac{q-1}{w_1 w_2} + 1$.
\end{cor}

%
%
\section{Conclusion and Further Works}

The upper bounds for restricted separating codes as well as separating Reed-Solomon
codes and their optimality were dealt with in the paper. Developing upper bounds for separating
codes is still an important topic in theory and practice. \par
Restricted separation property is quite strong condition, thus it is assumed that the upper bound for
them will be still smaller than the one of simple separating codes. 
Therefore, improvement of Theorem 4 could be a possible topic. \par
From the work of \cite{fer} to this paper, we confirmed that Silverberg's conjecture is true in many
cases and it derives the optimal upper bound of separating Reed-Solomon codes. Experimental results
tell us that almost all (about 90 percent) Reed-Solomon codes except few cases with $w$ in 2-25 and
$q$ in 2-4096 meets the optimal bound $M \leq q^{\lceil \frac{n}{w_1 w_2} \rceil}$.
In-depth study on separating codes and algebraic geometry codes seems to allow the complete solution
to Silverberg's open problem.

\end{document}